\documentclass[pre,twocolumn,superscriptaddress,floatfix,showpacs,showkeys]{revtex4}

\usepackage[latin2]{inputenc}
\usepackage{amsmath}
\usepackage{amssymb}
\usepackage{epsfig}

\begin{document}

\title{Shear Zones in granular materials: Optimization in a
self-organized random potential}

\date{\today; version 0.5}

\author{J. T\"or\"ok}
\affiliation{Department of Chemical Information Technology, Budapest
 University of 
 Technology and Economics, H-1111 Budapest, Hungary}
\affiliation{Department of Theoretical Physics, Budapest University of 
 Technology and Economics, H-1111 Budapest, Hungary}
\author{T. Unger}
\affiliation{Department of Theoretical Physics, Budapest University of 
 Technology and Economics, H-1111 Budapest, Hungary}
\author{J. Kert\'esz}
\affiliation{Department of Theoretical Physics, Budapest University of 
 Technology and Economics, H-1111 Budapest, Hungary}
\author{D.E. Wolf}
\affiliation{Department of Physics, University Duisburg-Essen,
 D-47048 Duisburg, Germany}

\begin{abstract}
We introduce a model to describe the wide shear zones observed in
modified Couette cell experiments with granular material. The model is
a generalization of the recently proposed approach based on a
variational principle. The instantaneous shear band is identified with
the surface that minimizes the dissipation in a random potential that is
biased by the local velocity difference and pressure. The apparent
shear zone is the ensemble average of the instantaneous shear
bands. The numerical simulation of this model matches excellently with
experiments and has measurable predictions.
\end{abstract} 



\maketitle

\begin{figure}[t]
\epsfig{figure=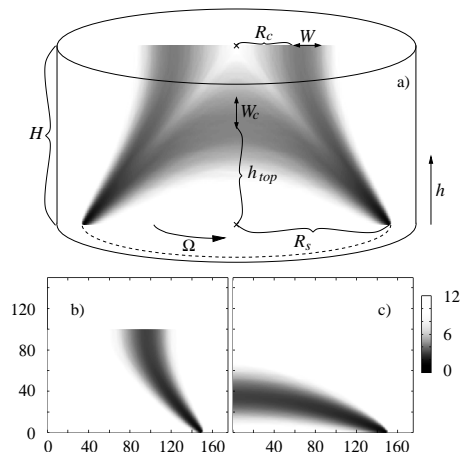,width=6cm}
\caption{
\label{Fig_setup}
a) The experimental arrangement and caluclated shear bands. The dotted
circle is the bottom split, the definition of the most important
quantities are noted on the figure. The gray scaling on all plots is
proportional to the logarithm of the occurrence probability
at the given point. $R_s=150$ for a-c) and $H=115,~ 100,~ 150$ for
plots a, b, c) respectively.  }
\end{figure}

Strain localization or shear band formation in granular materials has
been studied for many years \cite{DryGranu98,Jaeger96} due to its
importance in engineering and geoscience. {\it Shear bands} are narrow
regions separating almost solid blocks moving with different
velocities.  Recently Fenistein and coworkers observed
\cite{Fenistein2003a,Fenistein2004} wide {\it shear zones} in a
modified cylindrical Couette cell.  Instead of letting an inner
cylinder rotate with respect to an outer one, their cell has no inner
wall, but exerts the shear deformation via the bottom, which is split
into a rotating inner disk of radius $R_s$ and a fixed outer annulus
(see Fig. \ref{Fig_setup}). The shear zone is pinned at the bottom
split and evolves independently of walls. This approach has attracted
considerable interest \cite{Unger04,condmatNL,condmatUS} because it
provides new insight into the fundamental problem of shear band
formation.

A theory based on the principle of minimum dissipation rate was
proposed soon after the first publications. With the assumption of
negligible width of the shear zone a model with no fitting parameter
resulted which was used to describe the position of the shear zones
\cite{Unger04}. This model proved to be efficient and even delivered
predictions about a new type of closed shear zones, which have been
found later both in experiments and in computer simulations
\cite{condmatNL,condmatUS}.  However, the model has to be generalized
in order to describe the interesting phenomena related to the width of
the shear zone, and this is the aim of the present Letter.

Let us first briefly summarize the experimental findings: The shear
zone starts from the bottom split, and for small and moderate filling
$H\lesssim 0.7R_s$ it ends up on the surface \cite{Fenistein2003a}. If
the filling height is further increased, the shear zone is buried in
the material and takes the shape of a cupola
\cite{Unger04,condmatNL,condmatUS}. The surface position of the shear
zone for small filling can be very well described by a universal
empirical curve. The width of the shear zones on the surface increases
as a power law with the filling height with an exponent $\sim2/3$. For
the shape and width of closed shear zones, however, there are only
qualitative experimental results so far.

Our model combines the ideas of two main sources. The
first one is our former theoretical analysis of shear zones based on
the principle of minimum dissipation \cite{Unger04}. That model
(called ``Unger-model'' for short) assumes that the shear
zone is infinitely narrow, separating the standing and moving
parts of the material. The local dissipation rate is proportional
to the velocity difference across the shear band and to the hydrostatic pressure. In this
simple model the position of the shear band is identified with the shape
that corresponds to the global minimum dissipation rate. By
definition, this model was only able to describe the position of the
shear band.  Due to the cylindrical symmetry the problem was traced
back to finding the minimum path in a smooth 
2D potential. In spite of its simplicity, the model gave surprisingly
accurate results, moreover, it predicted the transition from the open
(Fig. \ref{Fig_setup} b) to the closed shapes (Fig. \ref{Fig_setup} c).

Another optimization problem was introduced earlier in the context of
shear band localization during compactification in sheared loose
granular matter \cite{Torok00}. In that model an instantaneous shear
bands were identified by the global minimum of a random potential
representing the local inhogeneities of the granular material. These
inhogeneities are not a priori frozen in but change due to the
relative displacement in the shear band. Thus the global minimum may
be at a different place in the next moment giving rise to an ansamble
of instantaneous shear bands which can be identified to the visible
macroscopic displacement field. These two models are here combined,
and the fluctuations induced by the random potential are used to
address the problem of the width of the shear bands.

We generalize the Unger-model in the following way: Instead of using a
smoothly varying potential like in \cite{Unger04} we use a random one
which is suited better to the disordered nature of granular
media. Like in ref. \cite{Torok00}, the instantaneous
displacement is represented by a single localized shear band
determined by optimization on a random field: The shear band is the
path which minimizes the dissipation rate and fits the boundary
conditions. (We shall again assume that, due to the symmetry of the
problem, the determination of the band reduces to that of a line.) The
shear is known to change the local structure of the material which we
take into account by changing the randomness in the neighborhood of
the actual shear band. A new optimal shear band is then searched
for. For each random realization of the material the minimization
determines a single narrow band. The shear zone itself is represented
as an ensemble of narrow bands, i.e. the flow velocity of the material
can be obtained as an ensemble average over the realizations. Our
approach is related to the first passage percolation problem
\cite{havlin00}, also known as polymer in a random
medium\cite{zhang95}, with the difference that in our case the
randomness organizes itself dynamically.

The model is defined on a regular square lattice which is applied as a
coarse grained representation of the material in a radial cut from the
center to the outer wall.
In the radial cut the shear band
is re\-presented by a continuous path that starts from the split point
of the bottom and reaches either the surface or the axis of the
sample. We allow nearest and next nearest neighbour connections
with lengths $\Delta \ell$ equal to $1$ and $\sqrt{2}$
respectively. Such a path ${\cal P}$ stands for a possible sliding
surface with cylindrical symmetry.

The energy dissipation rate associated with path ${\cal P}$ is in this
geometry proportional to the torque due to the local friction forces.
These are modelled by a random strength parameter $u(r,h)$ assigned to each
site of the lattice, where $r$ and $h$ are the radial and height
coordinates. Each random
variable $u$ is generated uniformly in $[\alpha u_{max},u_{max}]$,
$0\leq\alpha \leq 1$.  As in Ref.\cite{Unger04} we set $u_{max} = r^2
(H-h)$, which up to constant factors is hydrostatic pressure ($H-h$)
times cylindrical circumference ($2 \pi r$) times lever ($r$). The torque is
then given by integrating the local 
shear resistance of the material over the path ${\cal P}$:
${\cal S}=\sum_{i\in\cal P} u_i \Delta \ell_i$. 
The actual instantaneous shear band is the directed path
\cite{Kertesz04} that can be activated
by the smallest torque, i.e. the one which minimizes ${\cal S}$.  

Once the minimal path is found we refresh the strength parameters $u$
randomly along it and in its vicinity (nearest neighbour
sites). By the successive application of this procedure an ensemble of
shear bands is collected which provides the velocity field of the
shear flow. One instantaneous shear band separates the sample into two
parts in such a way that each site in the inner part rotates by the
driving angular velocity $\Omega$ of the bottom disk while the sites
in the outer part have angular velocity zero. Taking an average over
the ensemble of shear bands one arrives at a field of angular velocity
that can be compared directly to that observed in experiments.

An important difference between the model of \cite{Torok00} and the
present one is that the randomness is refreshed not only on the
current minimal path but also in its neighborhood. If only the values
in the path were changed no steady state would be reached. The average
value of the randomness $u$ would increase continuously, however
extremely slowly. This is not the case here where the steady state is
reached fast as in experiments. This variant of the model is closely
related to the self-organized criticality model of Bak and Sneppen
\cite{BakSneppen}.

The model has three free parameters. In dimensionless quantities they are:
$R_s/H$, $R_s/a$, $\alpha$. In words: the aspect ratio of the sample,
the split radius in units of the lattice constant and a number, which
controls the effect of disorder in the model. Henceforward in this paper
we use $a{=}1$, i.e. all lengths are measured in units of $a$.
The parameter $\alpha$ mimics strength
fluctuations due to individual properties (shape, friction, etc.) and
cooperative effects (e.g., density fluctuations). A smaller $\alpha$
means stronger disorder, while $\alpha=1$ would be a lattice version
of the deterministic Unger-model\cite{alpha}. We present data for
several different values $\alpha=0,~ 0.43,~ 0.5,~ 0.6$ and many
different split radii in the range of $R_s=15\dots 600$. As the
lattice unit must be larger than the lower length cutoff provided by
the particle diameter, larger values of $R_s$ correspond to smaller
grain size. 

In Fig. \ref{Fig_setup} the probability distribution of
instantaneous shear band positions is plotted. We get very similar
patterns as the ones
obtained in experiments and molecular dynamics simulations
\cite{condmatNL,condmatUS} with both open and closed shear zones.  We
calculate the angular velocity at any point of the sample and compare it
to the experiments.

If the shear zones are far from the system boundaries the error
function is a very good fit for the angular velocity in agreement with the
experiments \cite{Fenistein2003a}. It gives both position and width of
the zones.

\begin{figure}[t]
\epsfig{figure=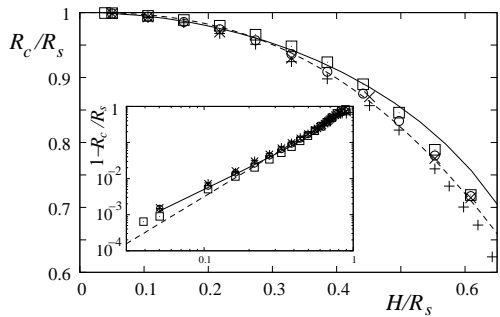,width=6.5cm}
\caption{
\label{Fig_surpos}
The surface positions of the shear zones. The solid line is from
\cite{Unger04}, the dashed line is the experimental
\cite{Fenistein2003a}
curve. Symbols were obtained for systems with $R_s=90$ and $\alpha=0,~
0.43,~ 0.5,~ 0.6$ for $+$, $\times$, $\circ$, $\boxdot$,
respectively. The inset shows the same data on log-log plot.  }
\end{figure}

Most of the experimental data concern the surface position of the
shear zones ($R_c$). We compare first
this property on Fig. \ref{Fig_surpos}. The analytical result of
\cite{Unger04} overestimated the experiments for $H/R_s\gtrsim 0.25$.
Increasing randomness decreases the apparent shear zone radius with
the best matching at about $\alpha\simeq 0.5$. System size does not
influence the curves for $R_s\gg10$.

From the experiments, Fenistein {\it et al.} \cite{Fenistein2003a}
suggested a power law dependence of the surface position on the
height, namely $1-R_c/R_s=(H/R_s)^{2.5}$. Thus on the inset of
Fig. \ref{Fig_surpos} we tested it on a log-log plot. Both the
Unger-model calculation and our
numerical data deviate slightly from the simple power law
function for small $H/R_s$. This deviation is too small to be seen on
a normal plot but could be tested on the experiments if high enough
precision can be attained \cite{finite size}.

\begin{figure}[t]
\epsfig{figure=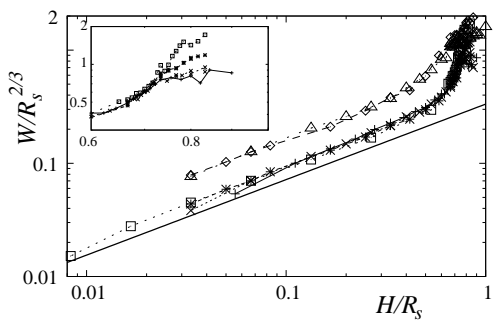,width=6.5cm}
\caption{
\label{Fig_surwidth}
The surface width of the shear zones.
The solid line is $(H/R_s)^{2/3}$. The lower curves are for
$\alpha=0.43$ with $R_s=90,~ 150,~ 300,~ 600$ with
symbols $+$, $\times$, $\ast$, $\boxdot$, respectively. The upper
curves are for $\alpha=0$ with $R_s=150,~ 300$ with symbols
$\diamond$, $\triangle$ respectively. The breakdown of scaling close
to the transition is shown in the inset for $\alpha=0.43$.
}
\end{figure}

In the experiments the width of the shear zones on the surface was
found to be a power law of $H$ with an exponent of $2/3$
\cite{condmatNL}.  Directed polymers have the same roughening
exponent\cite{zhang95}.  As shown on Fig. \ref{Fig_surwidth} we obtain
an exponent very close to this value.  The curves for different $R_s$
(but the same $\alpha$) can be scaled together by plotting
$W/R_s^{2/3}$ versus $H/R_s$ as in the experiments
\cite{condmatNL}. 

This power law increase of $W$ with $H$ must stop, when the width of
the shear zone reaches $R_c$, i.e. the available distance between the
container axis and the average shear band position at the surface. For
larger $H$ this finite size effect implies that $W\approx R_c \propto
R_s$ (for given $H/R_s$), which explains the sudden 
increase and the loss of data collapse of the scaled curves in Fig.
\ref{Fig_surwidth} at about $H/R_s\simeq0.7$. This feature can also be
observed on the experimental results \cite{condmatNL}. It cannot be
interpreted as a sign of an increasing characteristic length scale at the
transition from open to closed shear zones.

We find that with decreasing $\alpha$ (i.e. increasing disorder) the width
increases. This agrees with the experiments, where more irregular
particles produced wider shear zones \cite{Fenistein2003a}. If
$\alpha$ is varied, not 
only the surface width but also its position changes slightly.
This seems to be a secondary effect not yet observed experimentally. 
Our prediction is that the scaled surface position $R_c/R_s$ of the
shear zones is slightly larger for smoother particles than for rough
ones.

\begin{figure}[t]
\epsfig{figure=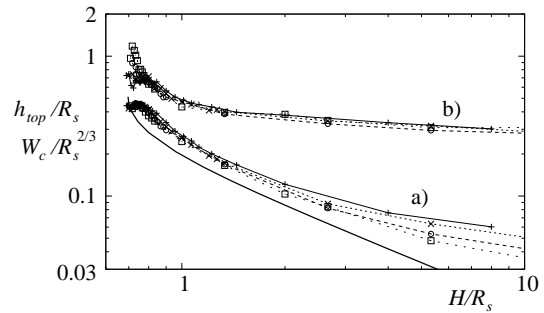,width=7cm}
\caption{
\label{Fig_cent}
The scaled center position (curves a) and width (curves b) of the
closed shear zones.  The solid line is the calculated position
from\cite{Unger04}.  Data\-points ($+$, $\times$, $\circ$, $\boxdot$)
connected with dashed
lines have $R_s{=}90,~ 150,~ 300,~ 600$ and
$\alpha{=}0.43$.
} \end{figure}

Due to the occurence of closed shear zones the local angular velocity
on the symmetry axis of the container depends on $h$, being equal to
$\Omega$ at the bottom and decreasing monotoneously towards the
surface. By fitting this dependence again by an error function, we
determine the position $h_{top}$ and vertical width $W_c$ of the closed
shear zones.  This works as well as on the surface provided the
shear zone is not too close to the boundaries.  This fitting procedure has
a much broader range of applicability than the one with half a
Gaussian which works well only for small systems $R_s\lesssim 30$ with
moderate $H\lesssim R_s$ \cite{condmatUS}.

Some of our measured datasets are plotted on Fig. \ref{Fig_cent}. The
closed shear zones become flatter with increasing $H$. The position
scales with $R_s$ and follows the curve calculated in \cite{Unger04}
for values of $H/R_s$ between 0.8 and 2. The deviations for larger
$H/R_s$ can be understood by noting that the height of the shear zone
decreases faster than the width with increasing $H$. Above a certain
filling height the shear zones touch the bottom of the container.
This raises the apparent position of the shear zones compared to the
noisless system of \cite{Unger04}.  

The vertical width $W_c$ of the closed shear zones scales with
$R_s^{2/3}$, as expected for directed polymers with a length of the
order $R_s$.  Close to the transition, for $H/R_s$ between 0.7 and
0.8, both $W_c$ and $h_{top}$ deviate from the expected
behaviour. Again this can be understood as a boundary effect: 
In the absence of fluctuations, $h_{top}$ is closer to the
surface than the width $W_c$ permits. 

\begin{figure}[t]
\epsfig{figure=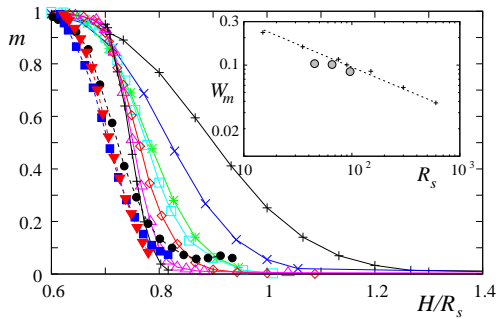,width=6.5cm}
\caption{
\label{Fig_order}
The order parameter $m$ for $\alpha=0$. System size $R_s=15,~ 35,~
75,~ 90,~ 150,~ 300,~ 600$ from right to left respectively ($+,~
\times,~ \ast,~ \boxdot,~ \diamond,~ \triangle,~ +$). Experimental
data from \cite{condmatNL} are shown with filled symbols of $\odot$,
$\bigtriangledown$, $\boxdot$ for $R_s=45,~ 65,~ 95$cm respectively.
The inset shows the width of the tranisition
versus $R_s$ for numerics $+$ and experiments $\odot$. The dashed line
has the slope of $0.5$.
}
\end{figure}

The next question we focus on is the phase transition. There seems to
be a discrepancy between the theoretical model \cite{Unger04} which
predicts a first order and the experiments \cite{condmatNL,condmatUS}
which claim to see a continuous transition. The latter two experimental
papers introduce different empirical fittings of the shear zone
profiles and show the transition of the fitting parameters. Both
papers estimate that the transition occurs at lower $H/R_S$
($H/R_S\simeq 0.6$ \cite{condmatUS}, $H/R_S\simeq 0.65$
\cite{condmatNL}) than in the theory of \cite{Unger04}. The drawback
of both approaches is that only one side of the transition can be
studied.

We prefer the classical approach of the order parameter of the
transition.  A good candidate seems to be the normalized angular velocity
of the surface at the center $m\equiv\omega(r{=}0,h{=}H)/\Omega$. If
the system has only open shear bands: $m=1$, if only closed ones:
$m=0$. If both types are present, $m$ can take any value
between 0 and 1. 

Fig. \ref{Fig_order} shows the change of $m$ with
$H/R_s$ for different $R_s$ and $\alpha=0$.
The transition gets sharper as the system size increases: 
In the thermodynamic limit $R_s \rightarrow \infty$, its
width seems to vanish like $R_s^{-1/2}$ (see inset of
Fig. \ref{Fig_order}). Then the order parameter 
jumps at a value of $H/R_s$, which we could estimate from a
finite size scaling analysis  as $(H/R_s)_c\simeq0.735$. 
Remarkably, this is very close to the higher limit of the
hysteresis calculated in \cite{Unger04}.

The angular velocity of the surface at the center is available in
experiments so that the test of the order parameter is straightforward. The
plot of $m$ for three systems can be found in Ref.  \cite{condmatNL}.
The experimental results look quite similar but with a little shift in
$H/R_s$. The sharpening of the transition in experiments cannot be
tested due to the limited range of $R_s$.

In conclusion we have shown that the results obtained from numerical
simulations of our model can be {\em directly} compared to the
experimental ones. Excellent agreement can be obtained for the already
measured quantities such as surface position, width and angular
velocity at the center of the surface. 
The comparison with experiments shows that our lattice constant can be    
indentified roughly with one particle diameter. This also
draws attention to the fact that the experimental systems
$R_s=15\dots95$ are far from the thermodynamic limit especially if one
studies the order of the transition. More attention should be given to
this fact.

The variational principle combined with the self-organized random
potential appears to be an efficient tool to study shear zones maybe
also in other geometries. The simplicity of the model with just a
single control parameter for the local strength fluctuations makes 
it robust, and applications to different geometries seem to be
straightforward.

We would like to thank to M.~van Hecke and J.B.~Lechman for useful
discussions. Partial support by grants OTKA F047259 and T049403, by
the G.I.F. research grant I-795-166.10/2003, as
well as by the Humboldt Foundation is acknowledged.


\begin{thebibliography}{99}
\bibitem{alpha} For $\alpha\gtrsim 0.7$ the lattice geometry has
strong influence on the shape of the shear bands. We restrict ourself
to the range of $\alpha\in[0:0.6]$.
\bibitem{finite size} We think that the deviations in the position of
the shear band from \cite{Unger04} and \cite{Fenistein2003a} reported
in \cite{condmatUS} for large $H/R_s\gtrsim0.6$ are due to the small
system size used by this group.
\end{thebibliography}
\end{document}